\documentclass[twocolumn,trackchanges]{aastex701}
\usepackage{color}

\newcommand{\Halpha}{\hbox{H$\alpha$}}
\newcommand{\Hbeta}{\hbox{H$\beta$}}

\newcommand{\HII}{\hbox{H\,{\sc ii}}}
\newcommand{\SIIdf}{\hbox{[S\,{\sc ii}]}$\lambda \lambda$ 4069,4076 }
\newcommand{\OIIdf}{\hbox{[O\,{\sc ii}]}$\lambda \lambda$ 7320,7330 }
\newcommand{\OIIIf}{\hbox{[O\,{\sc iii}]}$\lambda$ 4363 }
\newcommand{\SIIIf}{\hbox{[S\,{\sc iii}]}$\lambda$ 6312 }
\newcommand{\SIIId}{\hbox{[S\,{\sc iii}]}$\lambda \lambda$ 9069,9533 }

\newcommand{\NII}{\hbox{[N\,{\sc ii}]}$\lambda$ 6584 }
\newcommand{\NIIf}{\hbox{[N\,{\sc ii}]}$\lambda$ 5755 }
\newcommand{\SIId}{\hbox{[S\,{\sc ii}]}$\lambda \lambda$ 6716,6731 }
\newcommand{\OIId}{\hbox{[O\,{\sc ii}]}$\lambda \lambda$ 3726,3729 }
\newcommand{\OIII}{\hbox{[O\,{\sc iii}]}$\lambda$ 5007 }

\newcommand{\ha}{\hbox{H$\alpha$}}
\newcommand{\hb}{\hbox{H$\beta$}}
\newcommand{\oii}{\hbox{[O\,{\sc ii}]}}
\newcommand{\oiii}{\hbox{[O\,{\sc iii}]}}
\newcommand{\nii}{\hbox{[N\,{\sc ii}]}}
\newcommand{\sii}{\hbox{[S\,{\sc ii}]}}
\newcommand{\siii}{\hbox{[S\,{\sc iii}]}}
\newcommand{\caii}{\hbox{[Ca\,{\sc ii}]}}

\begin{document}

\title{Surprising Increase of Electron Temperature in Metal-Rich Star-Forming Regions}

\author[orcid=0009-0007-6697-8705,sname='Peng']{Ziming Peng}
\affiliation{Department of Physics, The Chinese University of Hong Kong, Shatin, New Territories, Hong Kong SAR, China}
\affiliation{JC STEM Lab of Astronomical Instrumentation, The Chinese University of Hong Kong, Shatin, New Territories, Hong Kong SAR, China}
\email[show]{zmpeng@link.cuhk.edu.hk}

\author[orcid=0000-0003-1025-1711,sname='Yan']{Renbin Yan} 
\affiliation{Department of Physics, The Chinese University of Hong Kong, Shatin, New Territories, Hong Kong SAR, China}
\affiliation{JC STEM Lab of Astronomical Instrumentation, The Chinese University of Hong Kong, Shatin, New Territories, Hong Kong SAR, China}
\affiliation{CUHK Shenzhen Research Institute, No.10, 2nd Yuexing Road, Nanshan, Shenzhen, China}
\email[show]{rbyan@cuhk.edu.hk}

\author[orcid=0000-0001-8078-3428,sname='Lin']{Zesen Lin}
\affiliation{Department of Physics, The Chinese University of Hong Kong, Shatin, New Territories, Hong Kong SAR, China}
\affiliation{JC STEM Lab of Astronomical Instrumentation, The Chinese University of Hong Kong, Shatin, New Territories, Hong Kong SAR, China}
\affiliation{CUHK Shenzhen Research Institute, No.10, 2nd Yuexing Road, Nanshan, Shenzhen, China}
\email{zslin@ustc.edu.cn}

\author[orcid=0000-0002-1660-9502,sname='Ji']{Xihan Ji}
\affiliation{Kavli Institute for Cosmology, University of Cambridge, Madingley Road, Cambridge CB3 0HA, UK}
\affiliation{Cavendish Laboratory, University of Cambridge, 19 JJ Thomson Avenue, Cambridge CB3 0HE, UK}
\email{xj274@cam.ac.uk}

\author[orcid=0009-0005-1799-5280,sname='Lee']{Man-Yin Leo Lee}
\affiliation{Department of Astronomy, University of California San Diego, 9500 Gilman Drive, La Jolla, CA 92093, USA}
\affiliation{Department of Physics, The Chinese University of Hong Kong, Shatin, New Territories, Hong Kong SAR, China}
\affiliation{JC STEM Lab of Astronomical Instrumentation, The Chinese University of Hong Kong, Shatin, New Territories, Hong Kong SAR, China}
\email{mal083@ucsd.edu}

\author[orcid=0000-0003-4520-5395,sname='chen']{Yuguang Chen}
\affiliation{Department of Physics, The Chinese University of Hong Kong, Shatin, New Territories, Hong Kong SAR, China}
\affiliation{JC STEM Lab of Astronomical Instrumentation, The Chinese University of Hong Kong, Shatin, New Territories, Hong Kong SAR, China}
\email{yuguangchen@cuhk.edu.hk}

\begin{abstract}

The electron temperature is a crucial parameter for the determination of the gas-phase metallicity of galaxies. Low electron temperature is expected for metal-rich galaxies, theoretically. We report the discovery that temperature, as measured through auroral-to-strong line ratios of O$^+$, trends in reverse directions at 12+log(O/H) $\geq$ 8.7. This trend remains consistent regardless of the emission line fitting method employed and is not attributable to contamination or dust attenuation correction. Notably, this phenomenon is not observed in other low-ionization ions, such as S$^+$ and N$^+$, which also probe electron temperature. The results are verified in two independent datasets. We analyze the potential cause for the high \oii\ auroral-to-strong line ratios at high metallicities, finding that no specific reason could account for that. This finding challenges the fundamental principles of the direct $T_e$ method for metallicity measurement, warranting further investigation into its physical interpretation.

\end{abstract}


\section{Introduction} 
\label{sec:1}

Measuring gas-phase metallicity, the relative abundance of elements heavier than helium to the abundance of hydrogen, is essential for constraining the chemical evolution of galaxies. Electron temperature plays a critical role as an intermediary in measuring gas-phase metallicity, as metallicity can be calculated from the ratio of the strong line flux of specific ion species (often oxygen) to the hydrogen recombination line flux (usually \hb) once the electron temperature is known \citep{draine2011physics}. Within the optical wavelength range, five sets of line ratios are commonly employed in optical spectroscopy to measure electron temperatures: \OIIdf / \OIId, \SIIdf / \SIId, \NIIf / \NII, \SIIIf / \SIIId, and \OIIIf / \OIII. In each set of line ratios, the numerator and denominator represent collisionally excited lines (CELs) of the same ion from different energy levels. The populations of the different energy levels have various dependencies on temperature \citep{1969BOTT....5....3P,peimbert2017nebular}. Thus, their ratio can yield a temperature measurement. Compared to the strong CELs (denominator), auroral lines (numerator) are usually two magnitudes fainter \citep{esteban2004reappraisal,berg2020chaos}. It is usually assumed that there are multiple ionization zones within a nebula, which can be probed by ions with different ionization potentials \cite{berg2015chaos}. In the optical band, $T_e$ (\oiii) is usually used to probe high ionization zones, while $T_e$ (\oii), $T_e$ (\nii), and $T_e$ (\sii) are used for understanding low ionization zones. Between them, the intermediate zones are usually represented by $T_e$ (\siii). 

In practical calculations of ionic abundance, the electron temperatures derived from the three low ionization species are often considered equivalent, with lower temperatures implying higher metallicity, as metal ions in high metallicity gas provide efficient cooling through collisional excitation and radiative de-excitation \citep{garnett1992electron}. However, this equality could not be true for supersolar metallicity star-forming galaxies and regions. Compared to $T_e$ (\nii) and $T_e$ (\sii), $T_e$ (\oii) tends to become much higher when the metallicity is high. In a previous paper \citep{peng2025dmd}, we discovered this systematic abnormal trend using Integrated Field Units (IFU) data. In this letter, we verify the relations using another independent single-fiber spectroscopy dataset. We discuss and analyze several causes for the upturn of $T_e$ (\oii), such as contamination, dust effects, and shock contribution. However, none of them can fully explain the observed phenomena.

This letter is structured as follows. Section \ref{sec:2} describes the data used in this work. The analysis processes we conduct to obtain electron temperatures are presented in Section \ref{sec:3}. Section \ref{sec:4} shows the results, and we discuss various potential reasons and conclude for these anomalies in Section \ref{sec:5}.

\section{Data}
\label{sec:2}
We select spectroscopic data for nearby galaxies that cover the wavelength range of auroral lines and their strong lines, including the Sloan Digital Sky Survey (hereafter SDSS) -IV Mapping the Nearby Galaxies at APO (hereafter MaNGA, \citealt{bundy2014overview,yan2016sdss}) and SDSS Legacy Survey (hereafter Legacy, \citealt{2000AJ....120.1579Y,abazajian2009seventh}). Both surveys can provide statistical measurements of electron temperatures using all three low-ionization zone indicators for a wide gas-phase metallicity range up to 12+log(O/H)=8.95. MaNGA is a spatially resolved IFU survey with $\sim 1-2 {\rm\ kpc}$ spatial resolution. On the other hand, Legacy targets each galaxy using a single fiber, covering the center region of $\sim 2-3 {\rm\ kpc}$ on average for each galaxy. 

From both surveys, we select star-forming regions using the criteria from \cite{ji2020constraining}, which provides a demarcation line based on strong line ratios (\NII / \Halpha, \SIId / \Halpha, and \OIII / \Hbeta). To detect \OIId and \OIIdf simultaneously and reverting them to the rest-frame, we set redshift limits for Legacy to 0.027 \textless $z$ \textless 0.25. We select MaNGA spaxels with $z$ \textless 0.08 following \cite{peng2025dmd}. With these selections, we have $\rm 1.5 \times 10^6$ spaxels from MaNGA and $\rm 2.7 \times 10^5$ fibers from Legacy for this study. 


\section{Deriving electron temperatures}
\label{sec:3}

To derive the theoretically calibrated strong line metallicity of each spectrum in Legacy and MaNGA, we utilize a photoionization model for star-forming regions \citep{ji2020constraining} generated by the photoionization code \textsc{Cloudy} v17.03 \citep{ferland20172017}. We assume an isobaric \HII\ region with plane-parallel geometry. The ionizing SEDs are constructed by \textsc{Starburst 99} v7.01 \citep{leitherer1999starburst99} with a continuous star-formation history and an age of 4 Myr and a Kroupa initial mass function (IMF, \citealt{kroupa2001variation}). The hydrogen density of this model is set to 14 cm$^{-3}$ which is derived from the median \sii$\lambda$ 6716/\sii$\lambda$ 6731 of \HII\ regions in MaNGA \citep{ji2020constraining}. We input the same gas-phase metallicities as the stellar metallicities, covering 0.05 to 3.16 $Z_\odot$, as the young massive stars likely have similar metallicities as their surrounding interstellar medium. The sulfur to oxygen ratio, S/O, is assumed to be the same as solar, but the relative abundances of carbon and nitrogen are assumed to increase with 12+log(O/H) following the relationship given by \cite{dopita2013new}. The ionization parameter, defined as $U = \frac{\phi_0}{n_H c}$, represents the relative strength of the ionizing radiation and is often denoted in logarithm ($log(U)$). This model prescription has been shown to closely reproduce the star-forming locus of MaNGA and Legacy samples in the 3D line ratio space composed of \NII / \Halpha, \SIId / \Halpha, and \OIII / \Hbeta. Using Bayesian inference \citep{blanc2015izi}, the joint and marginalized probability distribution functions (PDFs) of the strong line metallicity and ionization parameter can be found for each spectrum. We select the PDF-weighted averages as each spectrum's metallicity and ionization parameter, which are in good consistency with the most probable values \citep{ji2022correlation}. This metallicity calibration provides a good fit with the direct method, with a median discrepancy of 0.09 dex for MaNGA star-forming spaxels \citep{peng2025dmd}.

\begin{figure*}
    \centering
    \includegraphics[width=\linewidth]{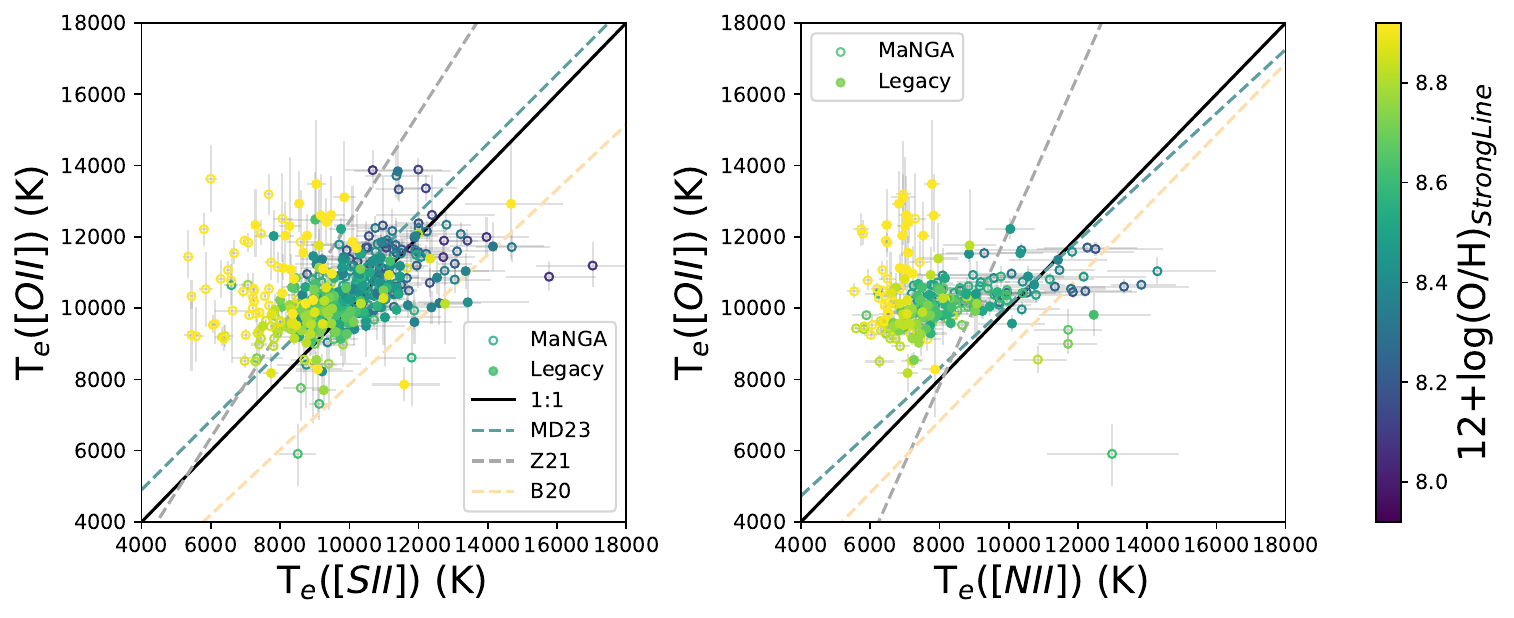}
    \caption{\textbf{Left:} Electron temperatures measured from \oii\ vs. those from \sii\ based on MaNGA () and Legacy () stacked spectra. \textbf{Right:} Electron temperatures measured from \oii\ vs. those from \nii\ based on MaNGA () and Legacy () stacked spectra. Each data point with error bars demonstrates a metallicity-ionization parameter bin and is color-coded by metallicity derived from strong line calibrations. The error bars correspond to the 1 $\sigma$ uncertainty of the temperature measurements. The black line shows the 1:1 line. The three dashed lines are the $T_e$ (\sii) versus $T_e$ (\oii) relation from \cite{mendez2023density,zurita2021bar,berg2020chaos}, respectively.}
    \label{fig:res}
\end{figure*}

After obtaining the metallicity and ionization parameter of each spectrum, we bin them into 0.05 $\times$ 0.05 dex boxes in these two parameters, and then stack the spectra within each bin. We only select bins that include more than 50 spectra to ensure high S/N ratios of stacked spectra. For the Legacy spectra, we also apply the correction provided by \cite{yan2011precision} to correct for the potential flux-calibration residuals. We measure emission line fluxes in the stacked spectra after the stellar continuum subtraction using the spectrum fitting code Penalized PiXel-Fitting (\textsc{pPXF}, \citealt{cappellari2017improving, Cappellari2023}) with stellar templates distilled from MaStar \citep{yan2019sdss} using a hierarchical clustering algorithm (MaStarHC, \citealt{westfall2019data}). For comparison, we also try another set of stellar templates from MILES \citep{2010MNRAS.404.1639V,2011A&A...532A..95F} to test the effect of the stellar continuum. The results imply that fittings with MaStarHC have statistically smaller reduced $\chi^2$. Our conclusions remain largely the same regardless of the choice of the stellar template set. 

Measuring the emission line flux is crucial in the data analysis. We select two sidebands and a central region for each emission line or doublet \citep{lee2024ionized}. A straight line is fitted using the residuals from two sidebands to model the local residuals of the stellar continuum for the emission line. We use Gaussian profiles to fit the emission line once the residual is subtracted, and use two Gaussian profiles with fixed wavelength separation for the doublets. The limited qualities of fittings for the continuum will result in incorrect line fluxes, especially for \SIIdf since the continuum near ${\rm H\delta}$ is difficult to subtract precisely. 

Dust extinction corrections could significantly impact the line ratio measurements. Typically, a whole spectrum is corrected by computing the Balmer decrement (\ha / \hb) to obtain the magnitude of extinction ($A_{\rm V}$) assuming an empirical extinction curve (e.g. \citealt{2019ApJ...886..108F}). However, the suitability of applying the same magnitude of extinction for all emission lines has been questioned \citep{ji2023need,2024A&A...691A.201L}, especially for data that do not spatially resolve individual \HII\ regions. It is suggested that emission lines from different ions may show different amounts of dust attenuation and thus cannot be corrected using a single $A_{\rm V}$ derived from the Balmer Decrement. This is likely because a single spectrum may contain mixtures of several \HII\ regions and diffuse ionized gas, and thus each ion may see a different flux-weighted average attenuation. We correct the dust attenuation following the method of \cite{ji2023need}. In each bin of fixed metallicity and ionization parameter, we divide the sample into sub-bins of different Balmer decrements and measure the auroral-to-strong line ratios in each sub-bin. We then fit the auroral-to-strong line ratios as a function of Balmer decrements across all sub-bins. This would yield an empirical correction for each ion. The result is consistent with \cite{ji2023need}, who show that the low ionization ions display less attenuation than that indicated by the Balmer lines. We utilize the empirical correction produced by this method to correct the auroral-to-strong line ratios for dust and calculate electron temperatures. 

Electron temperature and density are computed using {\it getCrossTemDen} from \textsc{PYNEB} \citep{luridiana2015pyneb,morisset2016photoionization}, with the default atomic data from PYNEB ('PYNEB\_23\_01'). We also checked the electron densities of the stacked spectra. Densities derived from \sii $\lambda$ 6716/\sii $\lambda$ 6731 of almost all the bins are under 200 cm$^{-3}$, indicating that the collisional excitation is the dominant process in the star-forming regions observed. We will discuss the effects of electron density measurements in detail in Section \ref{sec:5}. However, at this stage, we assume the auroral-to-strong line ratios depend exclusively on electron temperatures.

\begin{figure*}
    \centering
    \includegraphics[width=\linewidth]{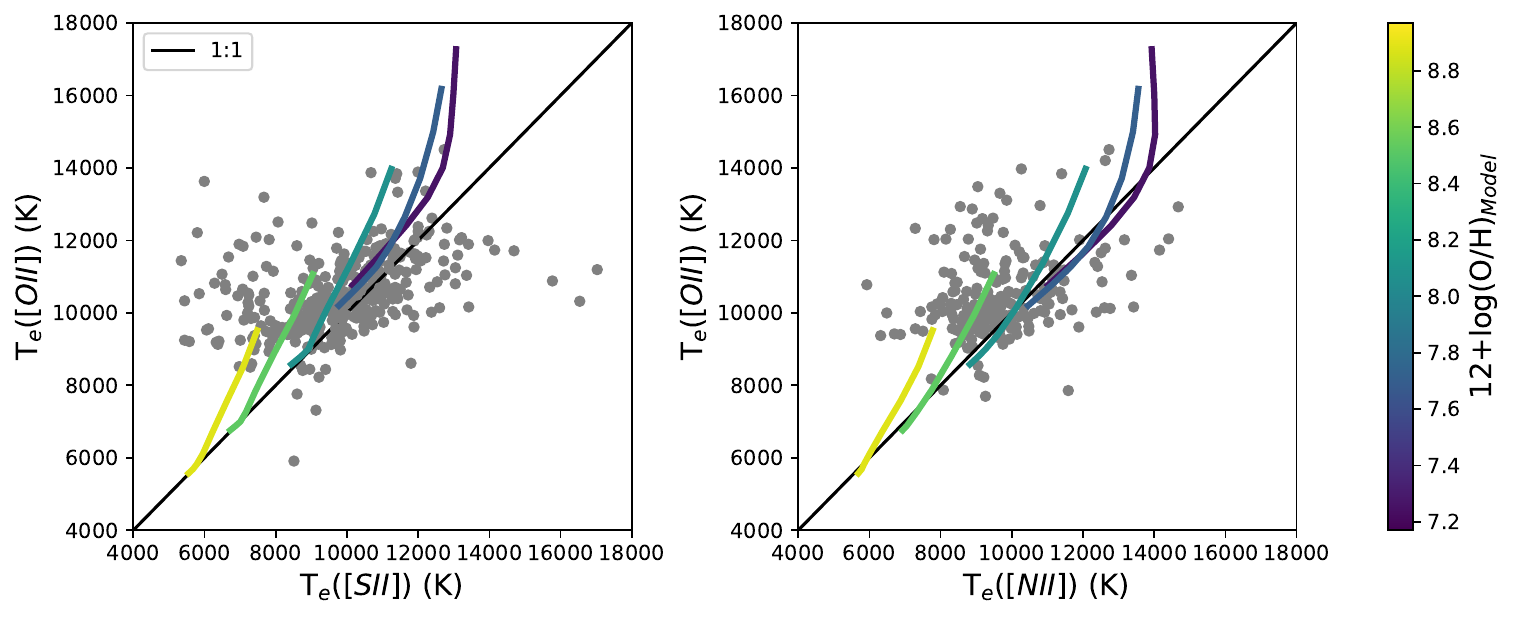}
    \caption{$T_e$ vs $T_e$ with the model considering the contributions of the recombination lines of both \OIIdf and \NIIf. The grey dots represent stacked observed data from MaNGA and Legacy. The colored lines are the temperatures derived from the model, color-coded by their gas-phase metallicities. The lines offset from the 1:1 line with the increasing of ionization parameters.}
    \label{fig:recomb}
\end{figure*}

\section{Results}
\label{sec:4}

We compare the electron temperatures derived from low ionization ions to assess their mutual consistency across different metallicity regimes. Our comparison of electron temperatures from \sii\ and \oii\ reveals an unexpected trend that deviates from the 1:1 line and the relations from previous studies \citep{mendez2023density,zurita2021bar,berg2020chaos}, as shown in the left panel of Fig. \ref{fig:res}. At low and intermediate metallicities, the data points are broadly distributed around the 1:1 line with moderate scatter, indicating overall consistency between the two electron temperature in these regimes. In agreement with theoretical expectations, both $T_e$ (\sii) and $T_e$ (\oii) decrease as metallicity increases, reflecting enhanced radiative cooling in more metal-rich gas. At high metallicities, however, we identify a clear departure from the 1:1 line. Stacked spectra of high metallicities, represented by green and yellow data points (shape in the fig), have increasing $T_e$ (\oii) while the metallicities become higher. Comparably, $T_e$ (\sii) still decrease with increasing metallicities in most bins. This behavior is already well observed in MaNGA samples \citep{peng2025dmd}, and here we verify that similar upturn is presented in the Legacy data, shown in filled circles. The consistency of these two surveys demonstrates that this abnormal upturn is not a data-reduction effect or survey-specific systematics. It represents the true results of measurements. 

We additionally examine the temperature derived from \NIIf / \NII versus $T_e$ (\oii) to validate this phenomenon. The \NIIf auroral lines lie at the junction of two spectrograph channels in both the Legacy and MaNGA surveys, leading to increased noise and larger residuals compared to auroral-to-strong line ratios from other ion species. As a result, $T_e$ (\nii) has large uncertainties. In the right panel of Fig .\ref{fig:res}, $T_e$ (\nii) - $T_e$ (\oii) exhibits fewer data points and large scatter at sub-solar metallicities. Nevertheless, they still show positive correlations. At high metallicities, $T_e$ (\nii) does not display increasing trend with increasing metallicity. This behavior mirrors that seen in the \sii–\oii\ comparison and further supports the conclusion that the increase in $T_e$ (\oii) at high metallicity is real. 

\begin{figure}
    
    \includegraphics[width=\linewidth]{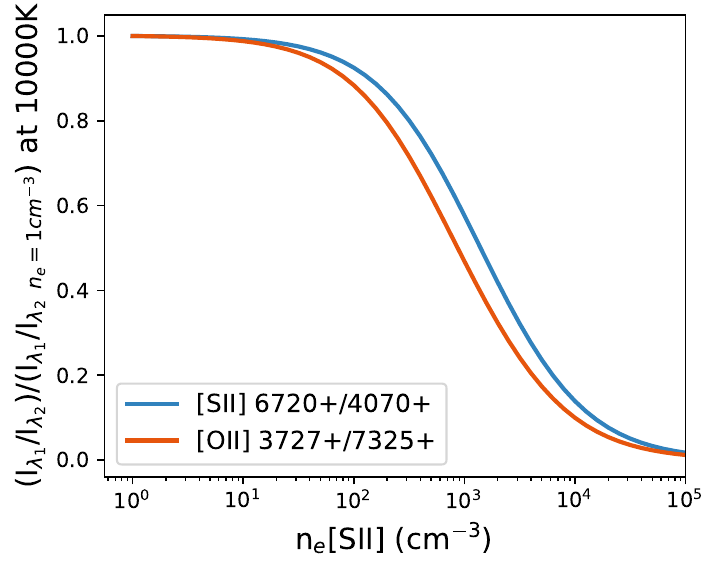}
    \caption{Variations of strong-to-auroral line ratios with different electron densities, normalized by the corresponding ratios at $n_e=1 {\rm cm}^{-3}$, assuming the electron temperature is 10,000K. 
    }
    \label{fig:density}
\end{figure}

\begin{figure*}
    \centering
    \includegraphics[width=\linewidth]{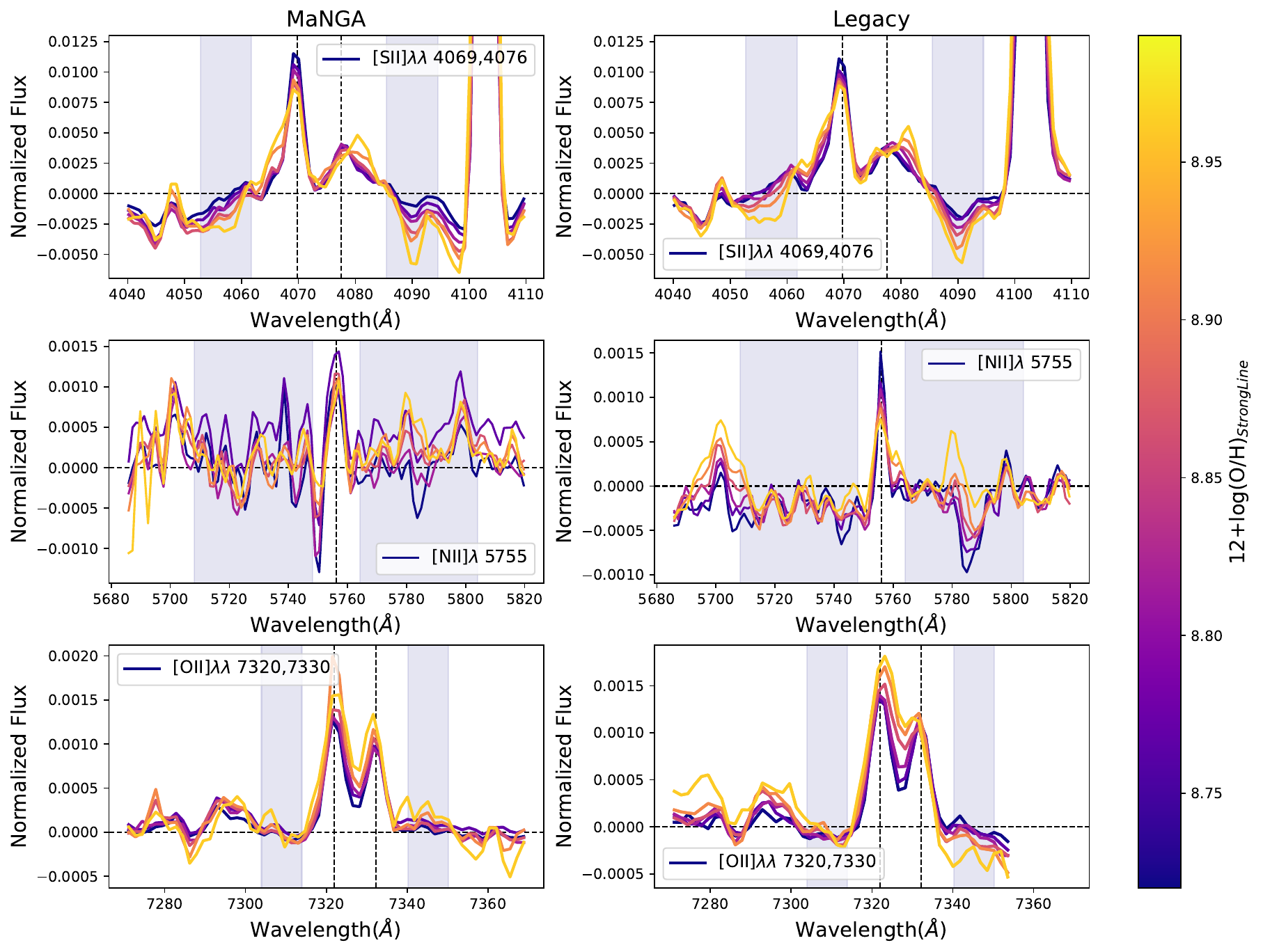}
    \caption{Part of the stacked spectra around \SIIdf (top), \NIIf (middle) and \OIIdf (bottom) after continuum subtraction. Here, the binning is only by the strong-line metallicity. Each spectrum is normalized by its strong line flux of the corresponding ion: the top panels are normalized by \SIId flux, the middle panels are normalized by \NII flux, and the bottom panels are normalized by \OIId flux. The shaded regions are the sidebands for fitting the auroral lines. All the spectra presented here are corrected for dust attenuation using the extinction curve from \cite{2019ApJ...886..108F}, assuming an intrinsic Balmer decrement \ha / \hb\ = 2.86.}
    \label{fig:spec}
\end{figure*}


\section{Reasons for temperature upturn}
\label{sec:5}

As the abnormal behavior of $T_e$ (\oii) is confirmed using two independent datasets, we aim to investigate the causes for this seemingly unphysical result. In this section, we discuss two groups of potential causes. First, we check for possibilities of contamination and other artifacts in the auroral line measurements, assuming that electron temperatures of low ionization species should be equal to each other in the ISM. Second, we discuss the potential mechanisms for raising $T_e$ (\oii), assuming that our measurements for emission lines are correct. At the end of this section, we compare similar results from previous works, and provide possible solutions that may further explain this issue. 

\subsection{Auroral line measurements}

We inspect the potential mechanisms that cause this abnormal increasing trend at high metallicity. It has been suggested that \oii\ auroral line measurements can be affected by imperfect telluric corrections \citep{yates2020present,2021MNRAS.502..225A}, which are corrections on certain wavelengths. However, we are selecting star-forming regions from a range of redshifts, the telluric contamination at a fixed wavelength in the observed spectrum would be spread over a wide window in the stacked spectra and hence has minimal impact. Another potential contaminating source is the \caii\ $\lambda$ 7323 emission line, which is blended together with the two \oii\ lines and provides higher fluxes than the actual \oii\ auroral line fluxes. We can exclude this possibility by the absence of detection of the other line in the \caii\ doublet, \caii\ $\lambda$ 7291, in all the stacked spectra. Therefore, telluric and Calcium contamination as the reason for the high auroral-to-strong line ratios are ruled out. 

One more possible explanation for the unrealistically high \oii\ auroral-to-strong line ratio is the contamination from non-collisionally excited lines or the dielectronic recombination lines near 7325 \AA\ \citep{1986ApJ...309..334R}, which is well-discussed in Planetary Nebulae (PNe) for \OIIIf \citep{2020MNRAS.498L..82G}. If the recombination contamination is considered, the photoionization model could predict an upturn of \OIIdf / \OIId when the electron temperature is low \citep{2005A&A...434..507S}. For PNe, the contamination fraction of recombination lines could be estimated using the method proposed by \cite{2000MNRAS.312..585L}. However, the above theory is based on the assumption that O$^{++}$ is the dominant ion of oxygen, or at least comparable with the O$^+$ abundance. For star-forming regions with 12+log(O/H) \textgreater 8.6, O$^{++}$ abundance is typically less than 10\% of O$^+$ abundance \citep{curti2017new,brazzini2024metallicity}. With the equation provided to calculate the recombination lines \citep{2000MNRAS.312..585L}, the contamination fraction will be smaller than 5\% in our sample. Nevertheless, we fully consider the contamination of the cascading recombination lines using photoionization models \textsc{Cloudy}, and the results are presented in Fig. \ref{fig:recomb}. Compared to $T_e$ (\sii) and $T_e$ (\nii), it is true that $T_e$ (\oii) is lifted up after considering the recombination lines. But the theoretical $T_e$ (\oii) predicted by the photoionization model are much higher than $T_e$ (\sii) and $T_e$ (\nii) at metal-poor regimes, and at high metallicities, the predicted electron temperature only reaches $\sim$ 11,000 K considering the effect of recombination lines, while some of the data reaches $\sim$ 14,000 K. So the contamination of recombination lines is insufficient to account for the increasing trend. 

Another mechanism that could possibly affect the electron temperature measurement is the density inhomogeneity, as the inclusion of high-density regions may result in the overestimation of electron temperature for \sii\ and \oii\ \citep{mendez2023density}. In Fig. \ref{fig:density}, we show the theoretical line ratio changes with the electron density, derived from \textsc{PyNeb}, assuming a constant electron temperature of 10,000K. If there are regions with electron density significantly higher than 100 $cm^{-3}$, then \oii\ and \sii\ auroral-to-strong line ratios could display strong density dependence, affecting their validity as a temperature indicator. If there are high density regions contributing significant fraction of auroral-line flux, then they could make the temperature measurement appear artificially high. However, this should affect both \oii\ and \sii\ auroral-to-strong line ratios by similar levels ---within 10\% of each other, which is inconsistent with the lack of a corresponding upturn in the $T_e$ (\sii). Therefore, density inhomogeneity does not provide a plausible explanation for the observations.

\begin{figure*}
    \centering
    \includegraphics[width=\linewidth]{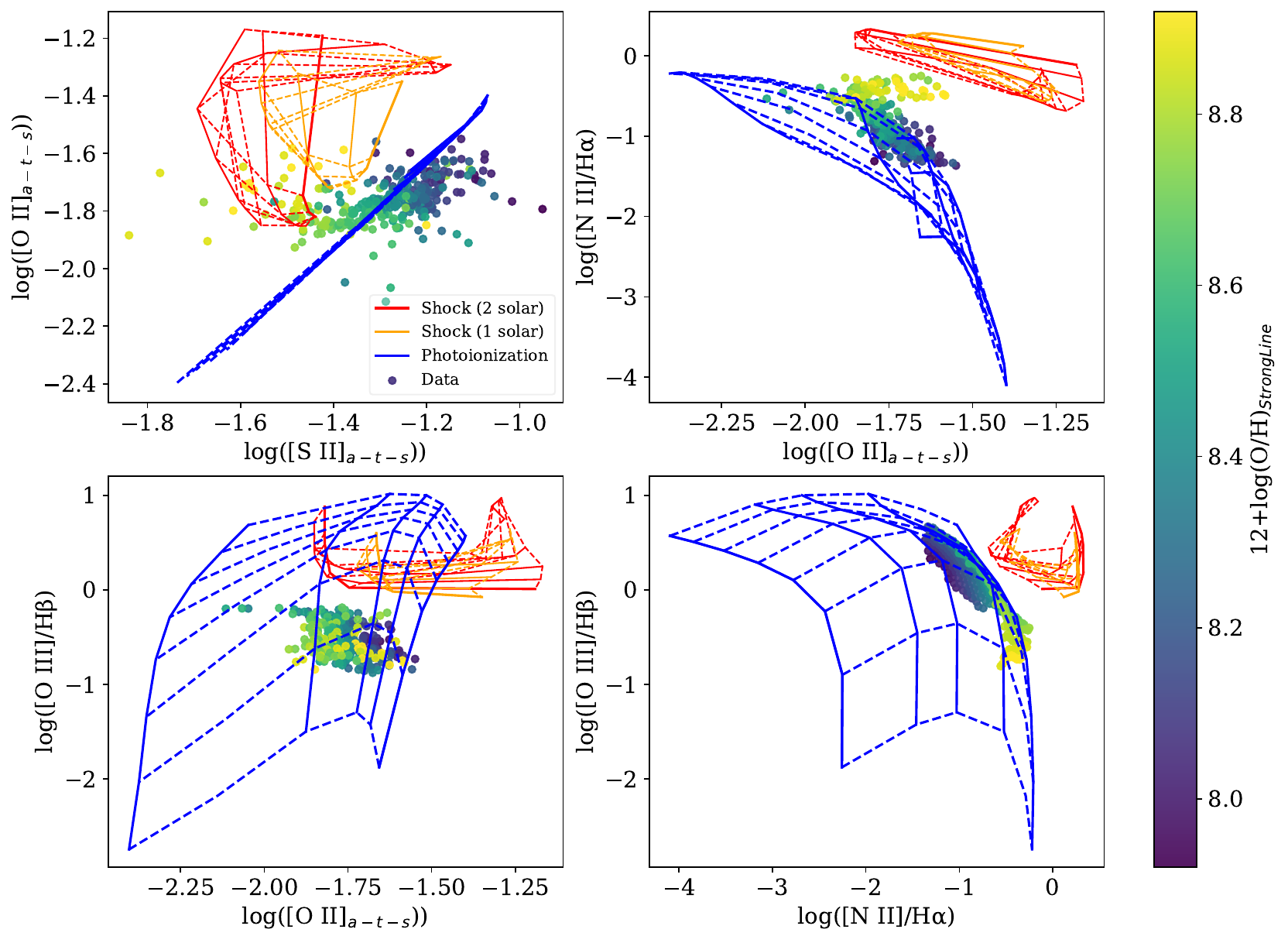}
    \caption{Line ratio comparisons between data and models. In each panel, the blue grids represent the photoionization model with different 12+log(O/H) and log(U), the red and orange grids represent the shock model with different B/n and velocity. Dots represent stacked data, color-coded by their strong-line metallicity. \oii$_{a-t-s}$ is \OIIdf/\OIId, \sii$_{a-t-s}$ is \SIIdf/\SIId.}
    \label{fig:shock}
\end{figure*}

We demonstrate the robustness of our auroral-to-strong line ratio measurements. We marginalize their dependence on the ionization parameter by re-binning the spectra only by their strong-line metallicities. We normalize the continuum-subtracted stacked spectra in each bin by the strong line flux (\sii, \nii, or \oii) and directly compare the spectra in the auroral line regions among multiple metallicity bins. The results are shown in Fig. \ref{fig:spec} color-coded by the strong-line metallicity. 
In the bottom panels, \oii\ auroral lines are clearly enhanced with increasing metallicity, both for Legacy and MaNGA. Since the de-redshifted spectra of Legacy stop at 7360 \AA, we select a narrower continuum width when fitting the auroral lines. In the top panels, \sii\ $\lambda$ 4069 auroral lines weaken with increasing metallicity, while the weaker \sii\ $\lambda$ 4076 remains at relatively the same level when the metallicity changes. 
We also show the strong-line normalized \NIIf spectra in the middle panels. Although they are noisy, they appear consistent with the results from \sii\ and do not show an abnormal temperature increase. 
All the spectra are corrected for dust using \cite{fitzpatrick1999correcting}. In fact, no matter whether we correct for dust or not, and no matter what extinction correction method we use, the trend of an upturn in the \oii\ auroral-to-strong line ratios at high metallicity persists.

Except for the reasons above, $\kappa$ distribution was proposed to solve the inconsistency of electron temperature measurement \citep{2012ApJ...752..148N}. It is unlikely that this explanation holds here because the timescale of the $\kappa$ distribution is too short for our statistically significant results \citep{2018ApJ...862...30D}. Ionizing structure inside \HII\ regions should not be the reason either, as the ionizing potential of $O^+$ is between that of $S^+$ and $N^+$, while neither of them shows the upturn. Extended red emission could affect the measurement of \OIIdf \citep{2020Ap&SS.365...58W}, but we are not able to quantify its impact.

\subsection{Extreme electron temperature}

We also investigate the potential physical process that could raising auroral-to-strong line ratios for \oii. Shock excitation was proposed to be a possible reason that causes the high $T_e$ (\oii) in \cite{2003ApJ...591..801K}. 
We use the photoionization model that we describe in Sec. \ref{sec:2} from \textsc{Cloudy} and the shock model from \textsc{Mappings V} from the 3MdBs database \citep{2019RMxAA..55..377A}. For the shock model, we select the solar metallicity grids and twice the solar metallicity grids, with a density of $\rm 1cm^{-3}$ and varying shock velocities and magnetic field parameters. Fig. \ref{fig:shock} illustrates the comparisons between data and all the models in different line ratio spaces. For the two auroral-to-strong line ratios presented in the upper left panel, indeed, some extreme combinations of the shock velocities and the magnetic field parameters can cover several data points. However, there are still data points to the left of the shock model grids. These outliers exhibit lower \SIIdf/\SIId\ ratios—typical of star-forming regions—and elevated \OIIdf/\OIId\ ratios. Considering the lower right panel of Fig. \ref{fig:shock}, the stacked spectra show no clue of being affected or dominated by shock. When comparing auroral-to-strong line ratios and strong line ratios, in the upper right panel, the high metallicity data tend to be located in regions that are poorly explained by photoionization models, while they are even less likely to be caused by shock. In the lower left panel, the data are located only in photoionization regions. Although our incomplete understanding of the complex physics of shocks and the relative immaturity of existing shock models, compared to photoionization codes, limit our ability to quantify their exact contribution, shock excitation appears to be insufficient to explain the auroral-to-strong line ratios and all the strong line ratios simultaneously.

\subsection{Discussions and conclusions}

We conclude that $T_e$ (\oii), or the \oii\ auroral-to-strong line ratio, shows an increasing trend with increasing metallicity when 12+log(O/H) $\geq$ 8.7. What makes this even more interesting is that only $T_e$ (\oii) shows this trend, but not $T_e$ (\nii) or $T_e$ (\sii). The abnormally high \oii\ temperatures can also be seen in the works by \cite{2003ApJ...591..801K,andrews2013mass,2015ApJ...808...42C}, where they were ignored or considered as outliers. The binning method used by \cite{andrews2013mass} is different from our method, which may result in the mixing of galaxies with different metallicities, making the trend harder to discern. In the sample of \HII\ regions presented by \cite{2003ApJ...591..801K}, the only supersolar \HII\ region, H1013, shows a much higher \oii\ temperature than some lower metallicity regions, consistent with our result. In \cite{2015ApJ...808...42C}, a clear $T_e$ (\oii) upturn was also shown, as in Fig. \ref{fig:res}, and they decided to discard $T_e$ (\oii) and use $T_e$ (\nii). Our stacking method probes the average trend among all \HII\ regions and is more convincing than what one could show with a few \HII\ regions.

Although the physical reason for this positive correlation is still under investigation, it challenges the validity of using the \oii\ based direct $T_e$ method for metallicity measurement in the high metallicity regime. Verified in two independent datasets with more than 1 million spectra, we demonstrate that the upturn of the \oii\ auroral-to-strong line ratios at high metallicities is an intrinsic property of the data.
This scenario also suggests that the current photoionization model is overly simplistic in describing the nebula ionized by young massive stars. We need to consider more realistic \HII\ region models, especially for metal-rich \HII\ regions. 

To validate this finding in the future, it is important to measure \NIIf / \NII reliably, as it is considered the least-scattered temperature indicator that is independent of density. Also, it is most widely used for low ionization zones owing to the closeness of wavelengths of the two lines \citep{2020MNRAS.497..672A,mendez2023temperature,vaught2024investigating}. Due to \NIIf falling close to the transition wavelength between the two spectral channels of SDSS, high-quality $T_e$ (\nii) cannot be obtained from Legacy or MaNGA. In this work, we observe that $T_e$ (\nii) does not present the upturn when metallicity is high. In the future, higher-quality measurements would help confirm this. Observations of individual resolved metal-rich \HII\ regions would also provide valuable insights to understand the upturn.

\begin{acknowledgments}
We acknowledge the grant support by the National Natural Science Foundation of China (NSFC; grant No. 12373008, 12425302), the support by the Research Grant Council of Hong Kong (Project No. 14302522, 14303123), by the Direct grant from the Faculty of Science of CUHK, and the support by the Hong Kong Jockey Club Charities Trust through the project, JC STEM Lab of Astronomical Instrumentation and Jockey Club Spectroscopy Survey System. RY acknowledges support by the Hong Kong Global STEM Scholar Scheme (GSP028). Z.S.L. acknowledges the support from Hong Kong Innovation and Technology Fund through the Research Talent Hub program (PiH/022/22GS). Y.C. is supported by the Direct Grant for Research (C0010-4053720) from the Faculty of Science, the Chinese University of Hong Kong. 

Funding for the Sloan Digital Sky 
Survey IV has been provided by the 
Alfred P. Sloan Foundation, the U.S. 
Department of Energy Office of 
Science, and the Participating 
Institutions. 

SDSS-IV acknowledges support and 
resources from the Center for High 
Performance Computing  at the 
University of Utah. The SDSS 
website is www.sdss4.org.

SDSS-IV is managed by the 
Astrophysical Research Consortium 
for the Participating Institutions 
of the SDSS Collaboration including 
the Brazilian Participation Group, 
the Carnegie Institution for Science, 
Carnegie Mellon University, Center for 
Astrophysics | Harvard \& 
Smithsonian, the Chilean Participation 
Group, the French Participation Group, 
Instituto de Astrof\'isica de 
Canarias, The Johns Hopkins 
University, Kavli Institute for the 
Physics and Mathematics of the 
Universe (IPMU) / University of 
Tokyo, the Korean Participation Group, 
Lawrence Berkeley National Laboratory, 
Leibniz Institut f\"ur Astrophysik 
Potsdam (AIP),  Max-Planck-Institut 
f\"ur Astronomie (MPIA Heidelberg), 
Max-Planck-Institut f\"ur 
Astrophysik (MPA Garching), 
Max-Planck-Institut f\"ur 
Extraterrestrische Physik (MPE), 
National Astronomical Observatories of 
China, New Mexico State University, 
New York University, University of 
Notre Dame, Observat\'ario 
Nacional / MCTI, The Ohio State 
University, Pennsylvania State 
University, Shanghai 
Astronomical Observatory, United 
Kingdom Participation Group, 
Universidad Nacional Aut\'onoma 
de M\'exico, University of Arizona, 
University of Colorado Boulder, 
University of Oxford, University of 
Portsmouth, University of Utah, 
University of Virginia, University 
of Washington, University of 
Wisconsin, Vanderbilt University, 
and Yale University.

Funding for the SDSS and SDSS-II has been provided by the Alfred P. Sloan Foundation, the Participating Institutions, the National Science Foundation, the U.S. Department of Energy, the National Aeronautics and Space Administration, the Japanese Monbukagakusho, the Max Planck Society, and the Higher Education Funding Council for England. The SDSS Web Site is http://www.sdss.org/.

The SDSS is managed by the Astrophysical Research Consortium for the Participating Institutions. The Participating Institutions are the American Museum of Natural History, Astrophysical Institute Potsdam, University of Basel, University of Cambridge, Case Western Reserve University, University of Chicago, Drexel University, Fermilab, the Institute for Advanced Study, the Japan Participation Group, Johns Hopkins University, the Joint Institute for Nuclear Astrophysics, the Kavli Institute for Particle Astrophysics and Cosmology, the Korean Scientist Group, the Chinese Academy of Sciences (LAMOST), Los Alamos National Laboratory, the Max-Planck-Institute for Astronomy (MPIA), the Max-Planck-Institute for Astrophysics (MPA), New Mexico State University, Ohio State University, University of Pittsburgh, University of Portsmouth, Princeton University, the United States Naval Observatory, and the University of Washington.
\end{acknowledgments}

\facilities{SDSS-IV/MaNGA, SDSS Legacy Survey}

\bibliography{all_ref}{}

@ARTICLE{2019RMxAA..55..377A,
       author = {{Alarie}, A. and {Morisset}, C.},
        title = "{Extensive Online Shock Model Database}",
      journal = {\rmxaa},
         year = 2019,
        month = oct,
       volume = {55},
        pages = {377-394},
          doi = {10.22201/ia.01851101p.2019.55.02.21},
archivePrefix = {arXiv},
       eprint = {1908.08579},
 primaryClass = {astro-ph.GA},
       adsurl = {https://ui.adsabs.harvard.edu/abs/2019RMxAA..55..377A}
}

@ARTICLE{2021MNRAS.502..225A,
       author = {{Arellano-C{\'o}rdova}, K.~Z. and {Esteban}, C. and {Garc{\'\i}a-Rojas}, J. and {M{\'e}ndez-Delgado}, J.~E.},
        title = "{On the radial abundance gradients of nitrogen and oxygen in the inner Galactic disc}",
      journal = {\mnras},
         year = 2021,
        month = mar,
       volume = {502},
       number = {1},
        pages = {225-241},
          doi = {10.1093/mnras/staa3903},
archivePrefix = {arXiv},
       eprint = {2012.06643},
 primaryClass = {astro-ph.GA},
       adsurl = {https://ui.adsabs.harvard.edu/abs/2021MNRAS.502..225A}
}

@ARTICLE{2020MNRAS.497..672A,
       author = {{Arellano-C{\'o}rdova}, K.~Z. and {Rodr{\'\i}guez}, M.},
        title = "{The T$_{e}$[N II]-T$_{e}$[O III] temperature relation in H II regions and the reliability of strong-line methods}",
      journal = {\mnras},
         year = 2020,
        month = sep,
       volume = {497},
       number = {1},
        pages = {672-686},
          doi = {10.1093/mnras/staa1759},
archivePrefix = {arXiv},
       eprint = {2006.07393},
 primaryClass = {astro-ph.GA},
       adsurl = {https://ui.adsabs.harvard.edu/abs/2020MNRAS.497..672A}
}

@ARTICLE{abazajian2009seventh,
       author = {{Abazajian}, Kevork N. and {Adelman-McCarthy}, Jennifer K. and {Ag{\"u}eros}, Marcel A. and {Allam}, Sahar S. and {Allende Prieto}, Carlos and {An}, Deokkeun and {Anderson}, Kurt S.~J. and {Anderson}, Scott F. and {Annis}, James and {Bahcall}, Neta A. and {Bailer-Jones}, C.~A.~L. and {Barentine}, J.~C. and {Bassett}, Bruce A. and {Becker}, Andrew C. and {Beers}, Timothy C. and {Bell}, Eric F. and {Belokurov}, Vasily and {Berlind}, Andreas A. and {Berman}, Eileen F. and {Bernardi}, Mariangela and {Bickerton}, Steven J. and {Bizyaev}, Dmitry and {Blakeslee}, John P. and {Blanton}, Michael R. and {Bochanski}, John J. and {Boroski}, William N. and {Brewington}, Howard J. and {Brinchmann}, Jarle and {Brinkmann}, J. and {Brunner}, Robert J. and {Budav{\'a}ri}, Tam{\'a}s and {Carey}, Larry N. and {Carliles}, Samuel and {Carr}, Michael A. and {Castander}, Francisco J. and {Cinabro}, David and {Connolly}, A.~J. and {Csabai}, Istv{\'a}n and {Cunha}, Carlos E. and {Czarapata}, Paul C. and {Davenport}, James R.~A. and {de Haas}, Ernst and {Dilday}, Ben and {Doi}, Mamoru and {Eisenstein}, Daniel J. and {Evans}, Michael L. and {Evans}, N.~W. and {Fan}, Xiaohui and {Friedman}, Scott D. and {Frieman}, Joshua A. and {Fukugita}, Masataka and {G{\"a}nsicke}, Boris T. and {Gates}, Evalyn and {Gillespie}, Bruce and {Gilmore}, G. and {Gonzalez}, Belinda and {Gonzalez}, Carlos F. and {Grebel}, Eva K. and {Gunn}, James E. and {Gy{\"o}ry}, Zsuzsanna and {Hall}, Patrick B. and {Harding}, Paul and {Harris}, Frederick H. and {Harvanek}, Michael and {Hawley}, Suzanne L. and {Hayes}, Jeffrey J.~E. and {Heckman}, Timothy M. and {Hendry}, John S. and {Hennessy}, Gregory S. and {Hindsley}, Robert B. and {Hoblitt}, J. and {Hogan}, Craig J. and {Hogg}, David W. and {Holtzman}, Jon A. and {Hyde}, Joseph B. and {Ichikawa}, Shin-ichi and {Ichikawa}, Takashi and {Im}, Myungshin and {Ivezi{\'c}}, {\v{Z}}eljko and {Jester}, Sebastian and {Jiang}, Linhua and {Johnson}, Jennifer A. and {Jorgensen}, Anders M. and {Juri{\'c}}, Mario and {Kent}, Stephen M. and {Kessler}, R. and {Kleinman}, S.~J. and {Knapp}, G.~R. and {Konishi}, Kohki and {Kron}, Richard G. and {Krzesinski}, Jurek and {Kuropatkin}, Nikolay and {Lampeitl}, Hubert and {Lebedeva}, Svetlana and {Lee}, Myung Gyoon and {Lee}, Young Sun and {French Leger}, R. and {L{\'e}pine}, S{\'e}bastien and {Li}, Nolan and {Lima}, Marcos and {Lin}, Huan and {Long}, Daniel C. and {Loomis}, Craig P. and {Loveday}, Jon and {Lupton}, Robert H. and {Magnier}, Eugene and {Malanushenko}, Olena and {Malanushenko}, Viktor and {Mandelbaum}, Rachel and {Margon}, Bruce and {Marriner}, John P. and {Mart{\'\i}nez-Delgado}, David and {Matsubara}, Takahiko and {McGehee}, Peregrine M. and {McKay}, Timothy A. and {Meiksin}, Avery and {Morrison}, Heather L. and {Mullally}, Fergal and {Munn}, Jeffrey A. and {Murphy}, Tara and {Nash}, Thomas and {Nebot}, Ada and {Neilsen}, Jr., Eric H. and {Newberg}, Heidi Jo and {Newman}, Peter R. and {Nichol}, Robert C. and {Nicinski}, Tom and {Nieto-Santisteban}, Maria and {Nitta}, Atsuko and {Okamura}, Sadanori and {Oravetz}, Daniel J. and {Ostriker}, Jeremiah P. and {Owen}, Russell and {Padmanabhan}, Nikhil and {Pan}, Kaike and {Park}, Changbom and {Pauls}, George and {Peoples}, Jr., John and {Percival}, Will J. and {Pier}, Jeffrey R. and {Pope}, Adrian C. and {Pourbaix}, Dimitri and {Price}, Paul A. and {Purger}, Norbert and {Quinn}, Thomas and {Raddick}, M. Jordan and {Re Fiorentin}, Paola and {Richards}, Gordon T. and {Richmond}, Michael W. and {Riess}, Adam G. and {Rix}, Hans-Walter and {Rockosi}, Constance M. and {Sako}, Masao and {Schlegel}, David J. and {Schneider}, Donald P. and {Scholz}, Ralf-Dieter and {Schreiber}, Matthias R. and {Schwope}, Axel D. and {Seljak}, Uro{\v{s}} and {Sesar}, Branimir and {Sheldon}, Erin and {Shimasaku}, Kazu and {Sibley}, Valena C. and {Simmons}, A.~E. and {Sivarani}, Thirupathi and {Allyn Smith}, J. and {Smith}, Martin C. and {Smol{\v{c}}i{\'c}}, Vernesa and {Snedden}, Stephanie A. and {Stebbins}, Albert and {Steinmetz}, Matthias and {Stoughton}, Chris and {Strauss}, Michael A. and {SubbaRao}, Mark and {Suto}, Yasushi and {Szalay}, Alexander S. and {Szapudi}, Istv{\'a}n and {Szkody}, Paula and {Tanaka}, Masayuki and {Tegmark}, Max and {Teodoro}, Luis F.~A. and {Thakar}, Aniruddha R. and {Tremonti}, Christy A. and {Tucker}, Douglas L. and {Uomoto}, Alan and {Vanden Berk}, Daniel E. and {Vandenberg}, Jan and {Vidrih}, S. and {Vogeley}, Michael S. and {Voges}, Wolfgang and {Vogt}, Nicole P. and {Wadadekar}, Yogesh and {Watters}, Shannon and {Weinberg}, David H. and {West}, Andrew A. and {White}, Simon D.~M. and {Wilhite}, Brian C. and {Wonders}, Alainna C. and {Yanny}, Brian and {Yocum}, D.~R.},
        title = "{The Seventh Data Release of the Sloan Digital Sky Survey}",
      journal = {\apjs},
         year = 2009,
        month = jun,
       volume = {182},
       number = {2},
        pages = {543-558},
          doi = {10.1088/0067-0049/182/2/543},
archivePrefix = {arXiv},
       eprint = {0812.0649},
 primaryClass = {astro-ph},
       adsurl = {https://ui.adsabs.harvard.edu/abs/2009ApJS..182..543A}
}

@ARTICLE{2015ApJ...808...42C,
       author = {{Croxall}, Kevin V. and {Pogge}, Richard W. and {Berg}, Danielle A. and {Skillman}, Evan D. and {Moustakas}, John},
        title = "{CHAOS II. Gas-phase Abundances in NGC 5194}",
      journal = {\apj},
         year = 2015,
        month = jul,
       volume = {808},
       number = {1},
          eid = {42},
        pages = {42},
          doi = {10.1088/0004-637X/808/1/42},
archivePrefix = {arXiv},
       eprint = {1501.02272},
 primaryClass = {astro-ph.GA},
       adsurl = {https://ui.adsabs.harvard.edu/abs/2015ApJ...808...42C}
}

@ARTICLE{2018ApJ...862...30D,
       author = {{Draine}, B.~T. and {Kreisch}, C.~D.},
        title = "{Electron Energy Distributions in H II Regions and Planetary Nebulae: {\ensuremath{\kappa}}-distributions Do Not Apply}",
      journal = {\apj},
         year = 2018,
        month = jul,
       volume = {862},
       number = {1},
          eid = {30},
        pages = {30},
          doi = {10.3847/1538-4357/aac891},
archivePrefix = {arXiv},
       eprint = {1803.10003},
 primaryClass = {astro-ph.GA},
       adsurl = {https://ui.adsabs.harvard.edu/abs/2018ApJ...862...30D}
}

@BOOK{draine2011physics,
       author = {{Draine}, Bruce T.},
        title = "{Physics of the Interstellar and Intergalactic Medium}",
         year = 2011,
       adsurl = {https://ui.adsabs.harvard.edu/abs/2011piim.book.....D},
    publisher = {Princeton University Press}
}

@ARTICLE{2019ApJ...886..108F,
       author = {{Fitzpatrick}, E.~L. and {Massa}, Derck and {Gordon}, Karl D. and {Bohlin}, Ralph and {Clayton}, Geoffrey C.},
        title = "{An Analysis of the Shapes of Interstellar Extinction Curves. VII. Milky Way Spectrophotometric Optical-through-ultraviolet Extinction and Its R-dependence}",
      journal = {\apj},
         year = 2019,
        month = dec,
       volume = {886},
       number = {2},
          eid = {108},
        pages = {108},
          doi = {10.3847/1538-4357/ab4c3a},
archivePrefix = {arXiv},
       eprint = {1910.08852},
 primaryClass = {astro-ph.GA},
       adsurl = {https://ui.adsabs.harvard.edu/abs/2019ApJ...886..108F}
}

@ARTICLE{2011A&A...532A..95F,
       author = {{Falc{\'o}n-Barroso}, J. and {S{\'a}nchez-Bl{\'a}zquez}, P. and {Vazdekis}, A. and {Ricciardelli}, E. and {Cardiel}, N. and {Cenarro}, A.~J. and {Gorgas}, J. and {Peletier}, R.~F.},
        title = "{An updated MILES stellar library and stellar population models}",
      journal = {\aap},
         year = 2011,
        month = aug,
       volume = {532},
          eid = {A95},
        pages = {A95},
          doi = {10.1051/0004-6361/201116842},
archivePrefix = {arXiv},
       eprint = {1107.2303},
 primaryClass = {astro-ph.CO},
       adsurl = {https://ui.adsabs.harvard.edu/abs/2011A&A...532A..95F}
}

@ARTICLE{2020MNRAS.498L..82G,
       author = {{G{\'o}mez-Llanos}, V. and {Morisset}, C. and {Garc{\'\i}a-Rojas}, J. and {Jones}, D. and {Wesson}, R. and {Corradi}, R.~L.~M. and {Boffin}, H.~M.~J.},
        title = "{The impact of strong recombination on temperature determination in planetary nebulae}",
      journal = {\mnras},
         year = 2020,
        month = nov,
       volume = {498},
       number = {1},
        pages = {L82-L86},
          doi = {10.1093/mnrasl/slaa131},
archivePrefix = {arXiv},
       eprint = {2007.05488},
 primaryClass = {astro-ph.GA},
       adsurl = {https://ui.adsabs.harvard.edu/abs/2020MNRAS.498L..82G}
}

@ARTICLE{2003ApJ...591..801K,
       author = {{Kennicutt}, Jr., Robert C. and {Bresolin}, Fabio and {Garnett}, Donald R.},
        title = "{The Composition Gradient in M101 Revisited. II. Electron Temperatures and Implications for the Nebular Abundance Scale}",
      journal = {\apj},
         year = 2003,
        month = jul,
       volume = {591},
       number = {2},
        pages = {801-820},
          doi = {10.1086/375398},
archivePrefix = {arXiv},
       eprint = {astro-ph/0303452},
 primaryClass = {astro-ph},
       adsurl = {https://ui.adsabs.harvard.edu/abs/2003ApJ...591..801K}
}

@ARTICLE{2000MNRAS.312..585L,
       author = {{Liu}, X. -W. and {Storey}, P.~J. and {Barlow}, M.~J. and {Danziger}, I.~J. and {Cohen}, M. and {Bryce}, M.},
        title = "{NGC 6153: a super-metal-rich planetary nebula?}",
      journal = {\mnras},
         year = 2000,
        month = mar,
       volume = {312},
       number = {3},
        pages = {585-628},
          doi = {10.1046/j.1365-8711.2000.03167.x},
       adsurl = {https://ui.adsabs.harvard.edu/abs/2000MNRAS.312..585L}
}

@ARTICLE{2024A&A...691A.201L,
       author = {{Lin}, Zesen and {Yan}, Renbin},
        title = "{Nebular dust attenuation with the Balmer and Paschen lines based on the MaNGA survey}",
      journal = {\aap},
         year = 2024,
        month = nov,
       volume = {691},
          eid = {A201},
        pages = {A201},
          doi = {10.1051/0004-6361/202451339},
archivePrefix = {arXiv},
       eprint = {2410.05067},
 primaryClass = {astro-ph.GA},
       adsurl = {https://ui.adsabs.harvard.edu/abs/2024A&A...691A.201L}
}

@ARTICLE{2012ApJ...752..148N,
       author = {{Nicholls}, David C. and {Dopita}, Michael A. and {Sutherland}, Ralph S.},
        title = "{Resolving the Electron Temperature Discrepancies in H II Regions and Planetary Nebulae: {\ensuremath{\kappa}}-distributed Electrons}",
      journal = {\apj},
         year = 2012,
        month = jun,
       volume = {752},
       number = {2},
          eid = {148},
        pages = {148},
          doi = {10.1088/0004-637X/752/2/148},
archivePrefix = {arXiv},
       eprint = {1204.3880},
 primaryClass = {astro-ph.GA},
       adsurl = {https://ui.adsabs.harvard.edu/abs/2012ApJ...752..148N}
}

@ARTICLE{1969BOTT....5....3P,
       author = {{Peimbert}, M. and {Costero}, R.},
        title = "{Chemical Abundances in Galactic HII Regions}",
      journal = {Boletin de los Observatorios Tonantzintla y Tacubaya},
         year = 1969,
        month = may,
       volume = {5},
        pages = {3-22},
       adsurl = {https://ui.adsabs.harvard.edu/abs/1969BOTT....5....3P}
}

@ARTICLE{peng2025dmd,
       author = {{Peng}, Ziming and {Yan}, Renbin and {Ji}, Xihan and {Lin}, Zesen and {Lee}, Man-Yin Leo},
        title = "{SDSS-IV MaNGA: Data-model discrepancy in temperature-sensitive line ratios for star-forming galaxies}",
      journal = {\aap},
         year = 2025,
        month = nov,
       volume = {703},
          eid = {A236},
        pages = {A236},
          doi = {10.1051/0004-6361/202556301},
archivePrefix = {arXiv},
       eprint = {2509.18770},
 primaryClass = {astro-ph.GA},
       adsurl = {https://ui.adsabs.harvard.edu/abs/2025A&A...703A.236P}
}

@ARTICLE{peimbert2017nebular,
       author = {{Peimbert}, Manuel and {Peimbert}, Antonio and {Delgado-Inglada}, Gloria},
        title = "{Nebular Spectroscopy: A Guide on Hii Regions and Planetary Nebulae}",
      journal = {\pasp},
         year = 2017,
        month = aug,
       volume = {129},
       number = {978},
        pages = {082001},
          doi = {10.1088/1538-3873/aa72c3},
archivePrefix = {arXiv},
       eprint = {1705.06323},
 primaryClass = {astro-ph.GA},
       adsurl = {https://ui.adsabs.harvard.edu/abs/2017PASP..129h2001P}
}

@ARTICLE{1986ApJ...309..334R,
       author = {{Rubin}, Robert H.},
        title = "{Noncollisional Excitation of Low-lying States in Gaseous Nebulae}",
      journal = {\apj},
         year = 1986,
        month = oct,
       volume = {309},
        pages = {334},
          doi = {10.1086/164606},
       adsurl = {https://ui.adsabs.harvard.edu/abs/1986ApJ...309..334R}
}

@ARTICLE{2005A&A...434..507S,
       author = {{Stasi{\'n}ska}, G.},
        title = "{Biases in abundance derivations for metal-rich nebulae}",
      journal = {\aap},
         year = 2005,
        month = may,
       volume = {434},
       number = {2},
        pages = {507-520},
          doi = {10.1051/0004-6361:20042216},
archivePrefix = {arXiv},
       eprint = {astro-ph/0501574},
 primaryClass = {astro-ph},
       adsurl = {https://ui.adsabs.harvard.edu/abs/2005A&A...434..507S}
}

@ARTICLE{2010MNRAS.404.1639V,
       author = {{Vazdekis}, A. and {S{\'a}nchez-Bl{\'a}zquez}, P. and {Falc{\'o}n-Barroso}, J. and {Cenarro}, A.~J. and {Beasley}, M.~A. and {Cardiel}, N. and {Gorgas}, J. and {Peletier}, R.~F.},
        title = "{Evolutionary stellar population synthesis with MILES - I. The base models and a new line index system}",
      journal = {\mnras},
         year = 2010,
        month = jun,
       volume = {404},
       number = {4},
        pages = {1639-1671},
          doi = {10.1111/j.1365-2966.2010.16407.x},
archivePrefix = {arXiv},
       eprint = {1004.4439},
 primaryClass = {astro-ph.CO},
       adsurl = {https://ui.adsabs.harvard.edu/abs/2010MNRAS.404.1639V}
}

@ARTICLE{2020Ap&SS.365...58W,
       author = {{Witt}, Adolf N. and {Lai}, Thomas S.-Y.},
        title = "{Extended red emission: observational constraints for models}",
      journal = {\apss},
         year = 2020,
        month = mar,
       volume = {365},
       number = {3},
          eid = {58},
        pages = {58},
          doi = {10.1007/s10509-020-03766-w},
archivePrefix = {arXiv},
       eprint = {2003.06453},
 primaryClass = {astro-ph.GA},
       adsurl = {https://ui.adsabs.harvard.edu/abs/2020Ap&SS.365...58W}
}

@ARTICLE{yan2011precision,
       author = {{Yan}, Renbin},
        title = "{Precision Spectrophotometry at the Level of 0.1\%}",
      journal = {\aj},
         year = 2011,
        month = nov,
       volume = {142},
       number = {5},
          eid = {153},
        pages = {153},
          doi = {10.1088/0004-6256/142/5/153},
archivePrefix = {arXiv},
       eprint = {1107.5040},
 primaryClass = {astro-ph.IM},
       adsurl = {https://ui.adsabs.harvard.edu/abs/2011AJ....142..153Y}
}

@ARTICLE{2000AJ....120.1579Y,
       author = {{York}, Donald G. and {Adelman}, J. and {Anderson}, Jr., John E. and {Anderson}, Scott F. and {Annis}, James and {Bahcall}, Neta A. and {Bakken}, J.~A. and {Barkhouser}, Robert and {Bastian}, Steven and {Berman}, Eileen and {Boroski}, William N. and {Bracker}, Steve and {Briegel}, Charlie and {Briggs}, John W. and {Brinkmann}, J. and {Brunner}, Robert and {Burles}, Scott and {Carey}, Larry and {Carr}, Michael A. and {Castander}, Francisco J. and {Chen}, Bing and {Colestock}, Patrick L. and {Connolly}, A.~J. and {Crocker}, J.~H. and {Csabai}, Istv{\'a}n and {Czarapata}, Paul C. and {Davis}, John Eric and {Doi}, Mamoru and {Dombeck}, Tom and {Eisenstein}, Daniel and {Ellman}, Nancy and {Elms}, Brian R. and {Evans}, Michael L. and {Fan}, Xiaohui and {Federwitz}, Glenn R. and {Fiscelli}, Larry and {Friedman}, Scott and {Frieman}, Joshua A. and {Fukugita}, Masataka and {Gillespie}, Bruce and {Gunn}, James E. and {Gurbani}, Vijay K. and {de Haas}, Ernst and {Haldeman}, Merle and {Harris}, Frederick H. and {Hayes}, J. and {Heckman}, Timothy M. and {Hennessy}, G.~S. and {Hindsley}, Robert B. and {Holm}, Scott and {Holmgren}, Donald J. and {Huang}, Chi-hao and {Hull}, Charles and {Husby}, Don and {Ichikawa}, Shin-Ichi and {Ichikawa}, Takashi and {Ivezi{\'c}}, {\v{Z}}eljko and {Kent}, Stephen and {Kim}, Rita S.~J. and {Kinney}, E. and {Klaene}, Mark and {Kleinman}, A.~N. and {Kleinman}, S. and {Knapp}, G.~R. and {Korienek}, John and {Kron}, Richard G. and {Kunszt}, Peter Z. and {Lamb}, D.~Q. and {Lee}, B. and {Leger}, R. French and {Limmongkol}, Siriluk and {Lindenmeyer}, Carl and {Long}, Daniel C. and {Loomis}, Craig and {Loveday}, Jon and {Lucinio}, Rich and {Lupton}, Robert H. and {MacKinnon}, Bryan and {Mannery}, Edward J. and {Mantsch}, P.~M. and {Margon}, Bruce and {McGehee}, Peregrine and {McKay}, Timothy A. and {Meiksin}, Avery and {Merelli}, Aronne and {Monet}, David G. and {Munn}, Jeffrey A. and {Narayanan}, Vijay K. and {Nash}, Thomas and {Neilsen}, Eric and {Neswold}, Rich and {Newberg}, Heidi Jo and {Nichol}, R.~C. and {Nicinski}, Tom and {Nonino}, Mario and {Okada}, Norio and {Okamura}, Sadanori and {Ostriker}, Jeremiah P. and {Owen}, Russell and {Pauls}, A. George and {Peoples}, John and {Peterson}, R.~L. and {Petravick}, Donald and {Pier}, Jeffrey R. and {Pope}, Adrian and {Pordes}, Ruth and {Prosapio}, Angela and {Rechenmacher}, Ron and {Quinn}, Thomas R. and {Richards}, Gordon T. and {Richmond}, Michael W. and {Rivetta}, Claudio H. and {Rockosi}, Constance M. and {Ruthmansdorfer}, Kurt and {Sandford}, Dale and {Schlegel}, David J. and {Schneider}, Donald P. and {Sekiguchi}, Maki and {Sergey}, Gary and {Shimasaku}, Kazuhiro and {Siegmund}, Walter A. and {Smee}, Stephen and {Smith}, J. Allyn and {Snedden}, S. and {Stone}, R. and {Stoughton}, Chris and {Strauss}, Michael A. and {Stubbs}, Christopher and {SubbaRao}, Mark and {Szalay}, Alexander S. and {Szapudi}, Istvan and {Szokoly}, Gyula P. and {Thakar}, Anirudda R. and {Tremonti}, Christy and {Tucker}, Douglas L. and {Uomoto}, Alan and {Vanden Berk}, Dan and {Vogeley}, Michael S. and {Waddell}, Patrick and {Wang}, Shu-i. and {Watanabe}, Masaru and {Weinberg}, David H. and {Yanny}, Brian and {Yasuda}, Naoki and {SDSS Collaboration}},
        title = "{The Sloan Digital Sky Survey: Technical Summary}",
      journal = {\aj},
         year = 2000,
        month = sep,
       volume = {120},
       number = {3},
        pages = {1579-1587},
          doi = {10.1086/301513},
archivePrefix = {arXiv},
       eprint = {astro-ph/0006396},
 primaryClass = {astro-ph},
       adsurl = {https://ui.adsabs.harvard.edu/abs/2000AJ....120.1579Y}
}

@ARTICLE{ji2020constraining,
       author = {{Ji}, Xihan and {Yan}, Renbin},
        title = "{Constraining photoionization models with a reprojected optical diagnostic diagram}",
      journal = {\mnras},
         year = 2020,
        month = dec,
       volume = {499},
       number = {4},
        pages = {5749-5764},
          doi = {10.1093/mnras/staa3259},
archivePrefix = {arXiv},
       eprint = {2007.09159},
 primaryClass = {astro-ph.GA},
       adsurl = {https://ui.adsabs.harvard.edu/abs/2020MNRAS.499.5749J}
}

@ARTICLE{bundy2014overview,
       author = {{Bundy}, Kevin and {Bershady}, Matthew A. and {Law}, David R. and {Yan}, Renbin and {Drory}, Niv and {MacDonald}, Nicholas and {Wake}, David A. and {Cherinka}, Brian and {S{\'a}nchez-Gallego}, Jos{\'e} R. and {Weijmans}, Anne-Marie and {Thomas}, Daniel and {Tremonti}, Christy and {Masters}, Karen and {Coccato}, Lodovico and {Diamond-Stanic}, Aleksandar M. and {Arag{\'o}n-Salamanca}, Alfonso and {Avila-Reese}, Vladimir and {Badenes}, Carles and {Falc{\'o}n-Barroso}, J{\'e}sus and {Belfiore}, Francesco and {Bizyaev}, Dmitry and {Blanc}, Guillermo A. and {Bland-Hawthorn}, Joss and {Blanton}, Michael R. and {Brownstein}, Joel R. and {Byler}, Nell and {Cappellari}, Michele and {Conroy}, Charlie and {Dutton}, Aaron A. and {Emsellem}, Eric and {Etherington}, James and {Frinchaboy}, Peter M. and {Fu}, Hai and {Gunn}, James E. and {Harding}, Paul and {Johnston}, Evelyn J. and {Kauffmann}, Guinevere and {Kinemuchi}, Karen and {Klaene}, Mark A. and {Knapen}, Johan H. and {Leauthaud}, Alexie and {Li}, Cheng and {Lin}, Lihwai and {Maiolino}, Roberto and {Malanushenko}, Viktor and {Malanushenko}, Elena and {Mao}, Shude and {Maraston}, Claudia and {McDermid}, Richard M. and {Merrifield}, Michael R. and {Nichol}, Robert C. and {Oravetz}, Daniel and {Pan}, Kaike and {Parejko}, John K. and {Sanchez}, Sebastian F. and {Schlegel}, David and {Simmons}, Audrey and {Steele}, Oliver and {Steinmetz}, Matthias and {Thanjavur}, Karun and {Thompson}, Benjamin A. and {Tinker}, Jeremy L. and {van den Bosch}, Remco C.~E. and {Westfall}, Kyle B. and {Wilkinson}, David and {Wright}, Shelley and {Xiao}, Ting and {Zhang}, Kai},
        title = "{Overview of the SDSS-IV MaNGA Survey: Mapping nearby Galaxies at Apache Point Observatory}",
      journal = {\apj},
         year = 2015,
        month = jan,
       volume = {798},
       number = {1},
          eid = {7},
        pages = {7},
          doi = {10.1088/0004-637X/798/1/7},
archivePrefix = {arXiv},
       eprint = {1412.1482},
 primaryClass = {astro-ph.GA},
       adsurl = {https://ui.adsabs.harvard.edu/abs/2015ApJ...798....7B}
}

@ARTICLE{yan2016sdss,
       author = {{Yan}, Renbin and {Bundy}, Kevin and {Law}, David R. and {Bershady}, Matthew A. and {Andrews}, Brett and {Cherinka}, Brian and {Diamond-Stanic}, Aleksandar M. and {Drory}, Niv and {MacDonald}, Nicholas and {S{\'a}nchez-Gallego}, Jos{\'e} R. and {Thomas}, Daniel and {Wake}, David A. and {Weijmans}, Anne-Marie and {Westfall}, Kyle B. and {Zhang}, Kai and {Arag{\'o}n-Salamanca}, Alfonso and {Belfiore}, Francesco and {Bizyaev}, Dmitry and {Blanc}, Guillermo A. and {Blanton}, Michael R. and {Brownstein}, Joel and {Cappellari}, Michele and {D'Souza}, Richard and {Emsellem}, Eric and {Fu}, Hai and {Gaulme}, Patrick and {Graham}, Mark T. and {Goddard}, Daniel and {Gunn}, James E. and {Harding}, Paul and {Jones}, Amy and {Kinemuchi}, Karen and {Li}, Cheng and {Li}, Hongyu and {Maiolino}, Roberto and {Mao}, Shude and {Maraston}, Claudia and {Masters}, Karen and {Merrifield}, Michael R. and {Oravetz}, Daniel and {Pan}, Kaike and {Parejko}, John K. and {Sanchez}, Sebastian F. and {Schlegel}, David and {Simmons}, Audrey and {Thanjavur}, Karun and {Tinker}, Jeremy and {Tremonti}, Christy and {van den Bosch}, Remco and {Zheng}, Zheng},
        title = "{SDSS-IV MaNGA IFS Galaxy Survey{\textemdash}Survey Design, Execution, and Initial Data Quality}",
      journal = {\aj},
         year = 2016,
        month = dec,
       volume = {152},
       number = {6},
          eid = {197},
        pages = {197},
          doi = {10.3847/0004-6256/152/6/197},
archivePrefix = {arXiv},
       eprint = {1607.08613},
 primaryClass = {astro-ph.GA},
       adsurl = {https://ui.adsabs.harvard.edu/abs/2016AJ....152..197Y}
}

@ARTICLE{westfall2019data,
       author = {{Westfall}, Kyle B. and {Cappellari}, Michele and {Bershady}, Matthew A. and {Bundy}, Kevin and {Belfiore}, Francesco and {Ji}, Xihan and {Law}, David R. and {Schaefer}, Adam and {Shetty}, Shravan and {Tremonti}, Christy A. and {Yan}, Renbin and {Andrews}, Brett H. and {Brownstein}, Joel R. and {Cherinka}, Brian and {Coccato}, Lodovico and {Drory}, Niv and {Maraston}, Claudia and {Parikh}, Taniya and {S{\'a}nchez-Gallego}, Jos{\'e} R. and {Thomas}, Daniel and {Weijmans}, Anne-Marie and {Barrera-Ballesteros}, Jorge and {Du}, Cheng and {Goddard}, Daniel and {Li}, Niu and {Masters}, Karen and {Ibarra Medel}, H{\'e}ctor Javier and {S{\'a}nchez}, Sebasti{\'a}n F. and {Yang}, Meng and {Zheng}, Zheng and {Zhou}, Shuang},
        title = "{The Data Analysis Pipeline for the SDSS-IV MaNGA IFU Galaxy Survey: Overview}",
      journal = {\aj},
         year = 2019,
        month = dec,
       volume = {158},
       number = {6},
          eid = {231},
        pages = {231},
          doi = {10.3847/1538-3881/ab44a2},
archivePrefix = {arXiv},
       eprint = {1901.00856},
 primaryClass = {astro-ph.GA},
       adsurl = {https://ui.adsabs.harvard.edu/abs/2019AJ....158..231W}
}

@ARTICLE{Cappellari2023,
       author = {{Cappellari}, Michele},
        title = "{Full spectrum fitting with photometry in PPXF: stellar population versus dynamical masses, non-parametric star formation history and metallicity for 3200 LEGA-C galaxies at redshift z {\ensuremath{\approx}} 0.8}",
      journal = {\mnras},
         year = 2023,
        month = dec,
       volume = {526},
       number = {3},
        pages = {3273-3300},
          doi = {10.1093/mnras/stad2597},
archivePrefix = {arXiv},
       eprint = {2208.14974},
 primaryClass = {astro-ph.GA},
       adsurl = {https://ui.adsabs.harvard.edu/abs/2023MNRAS.526.3273C}
}

@ARTICLE{fitzpatrick1999correcting,
       author = {{Fitzpatrick}, Edward L.},
        title = "{Correcting for the Effects of Interstellar Extinction}",
      journal = {\pasp},
         year = 1999,
        month = jan,
       volume = {111},
       number = {755},
        pages = {63-75},
          doi = {10.1086/316293},
archivePrefix = {arXiv},
       eprint = {astro-ph/9809387},
 primaryClass = {astro-ph},
       adsurl = {https://ui.adsabs.harvard.edu/abs/1999PASP..111...63F}
}

@ARTICLE{ji2023need,
       author = {{Ji}, Xihan and {Yan}, Renbin and {Bundy}, Kevin and {Boquien}, M{\'e}d{\'e}ric and {Schaefer}, Adam and {Belfiore}, Francesco and {Bershady}, Matthew A. and {Drory}, Niv and {Li}, Cheng and {Westfall}, Kyle B. and {Lin}, Zesen and {Bizyaev}, Dmitry and {Law}, David R. and {Riffel}, Rog{\'e}rio and {Riffel}, Rogemar A.},
        title = "{The need for multicomponent dust attenuation in modeling nebular emission: Constraints from SDSS-IV MaNGA}",
      journal = {\aap},
         year = 2023,
        month = feb,
       volume = {670},
          eid = {A125},
        pages = {A125},
          doi = {10.1051/0004-6361/202245072},
archivePrefix = {arXiv},
       eprint = {2209.13618},
 primaryClass = {astro-ph.GA},
       adsurl = {https://ui.adsabs.harvard.edu/abs/2023A&A...670A.125J}
}

@ARTICLE{ferland20172017,
       author = {{Ferland}, G.~J. and {Chatzikos}, M. and {Guzm{\'a}n}, F. and {Lykins}, M.~L. and {van Hoof}, P.~A.~M. and {Williams}, R.~J.~R. and {Abel}, N.~P. and {Badnell}, N.~R. and {Keenan}, F.~P. and {Porter}, R.~L. and {Stancil}, P.~C.},
        title = "{The 2017 Release Cloudy}",
      journal = {\rmxaa},
         year = 2017,
        month = oct,
       volume = {53},
        pages = {385-438},
          doi = {10.48550/arXiv.1705.10877},
archivePrefix = {arXiv},
       eprint = {1705.10877},
 primaryClass = {astro-ph.GA},
       adsurl = {https://ui.adsabs.harvard.edu/abs/2017RMxAA..53..385F}
}

@ARTICLE{leitherer1999starburst99,
       author = {{Leitherer}, Claus and {Schaerer}, Daniel and {Goldader}, Jeffrey D. and {Delgado}, Rosa M. Gonz{\'a}lez and {Robert}, Carmelle and {Kune}, Denis Foo and {de Mello}, Du{\'\i}lia F. and {Devost}, Daniel and {Heckman}, Timothy M.},
        title = "{Starburst99: Synthesis Models for Galaxies with Active Star Formation}",
      journal = {\apjs},
         year = 1999,
        month = jul,
       volume = {123},
       number = {1},
        pages = {3-40},
          doi = {10.1086/313233},
archivePrefix = {arXiv},
       eprint = {astro-ph/9902334},
 primaryClass = {astro-ph},
       adsurl = {https://ui.adsabs.harvard.edu/abs/1999ApJS..123....3L}
}

@ARTICLE{yan2019sdss,
       author = {{Yan}, Renbin and {Chen}, Yanping and {Lazarz}, Daniel and {Bizyaev}, Dmitry and {Maraston}, Claudia and {Stringfellow}, Guy S. and {McCarthy}, Kyle and {Meneses-Goytia}, Sofia and {Law}, David R. and {Thomas}, Daniel and {Falcon Barroso}, Jesus and {S{\'a}nchez-Gallego}, Jos{\'e} R. and {Schlafly}, Edward and {Zheng}, Zheng and {Argudo-Fern{\'a}ndez}, Maria and {Beaton}, Rachael L. and {Beers}, Timothy C. and {Bershady}, Matthew and {Blanton}, Michael R. and {Brownstein}, Joel and {Bundy}, Kevin and {Chambers}, Kenneth C. and {Cherinka}, Brian and {De Lee}, Nathan and {Drory}, Niv and {Galbany}, Llu{\'\i}s and {Holtzman}, Jon and {Imig}, Julie and {Kaiser}, Nick and {Kinemuchi}, Karen and {Liu}, Chao and {Luo}, A. -Li and {Magnier}, Eugene and {Majewski}, Steven and {Nair}, Preethi and {Oravetz}, Audrey and {Oravetz}, Daniel and {Pan}, Kaike and {Sobeck}, Jennifer and {Stassun}, Keivan and {Talbot}, Michael and {Tremonti}, Christy and {Waters}, Christopher and {Weijmans}, Anne-Marie and {Wilhelm}, Ronald and {Zasowski}, Gail and {Zhao}, Gang and {Zhao}, Yong-Heng},
        title = "{SDSS-IV MaStar: A Large and Comprehensive Empirical Stellar Spectral Library{\textemdash}First Release}",
      journal = {\apj},
         year = 2019,
        month = oct,
       volume = {883},
       number = {2},
          eid = {175},
        pages = {175},
          doi = {10.3847/1538-4357/ab3ebc},
archivePrefix = {arXiv},
       eprint = {1812.02745},
 primaryClass = {astro-ph.IM},
       adsurl = {https://ui.adsabs.harvard.edu/abs/2019ApJ...883..175Y}
}

@ARTICLE{kroupa2001variation,
       author = {{Kroupa}, Pavel},
        title = "{On the variation of the initial mass function}",
      journal = {\mnras},
         year = 2001,
        month = apr,
       volume = {322},
       number = {2},
        pages = {231-246},
          doi = {10.1046/j.1365-8711.2001.04022.x},
archivePrefix = {arXiv},
       eprint = {astro-ph/0009005},
 primaryClass = {astro-ph},
       adsurl = {https://ui.adsabs.harvard.edu/abs/2001MNRAS.322..231K}
}

@ARTICLE{dopita2013new,
       author = {{Dopita}, Michael A. and {Sutherland}, Ralph S. and {Nicholls}, David C. and {Kewley}, Lisa J. and {Vogt}, Fr{\'e}d{\'e}ric P.~A.},
        title = "{New Strong-line Abundance Diagnostics for H II Regions: Effects of {\ensuremath{\kappa}}-distributed Electron Energies and New Atomic Data}",
      journal = {\apjs},
         year = 2013,
        month = sep,
       volume = {208},
       number = {1},
          eid = {10},
        pages = {10},
          doi = {10.1088/0067-0049/208/1/10},
archivePrefix = {arXiv},
       eprint = {1307.5950},
 primaryClass = {astro-ph.CO},
       adsurl = {https://ui.adsabs.harvard.edu/abs/2013ApJS..208...10D}
}

@ARTICLE{cappellari2017improving,
       author = {{Cappellari}, Michele},
        title = "{Improving the full spectrum fitting method: accurate convolution with Gauss-Hermite functions}",
      journal = {\mnras},
         year = 2017,
        month = apr,
       volume = {466},
       number = {1},
        pages = {798-811},
          doi = {10.1093/mnras/stw3020},
archivePrefix = {arXiv},
       eprint = {1607.08538},
 primaryClass = {astro-ph.GA},
       adsurl = {https://ui.adsabs.harvard.edu/abs/2017MNRAS.466..798C}
}

@ARTICLE{curti2017new,
       author = {{Curti}, M. and {Cresci}, G. and {Mannucci}, F. and {Marconi}, A. and {Maiolino}, R. and {Esposito}, S.},
        title = "{New fully empirical calibrations of strong-line metallicity indicators in star-forming galaxies}",
      journal = {\mnras},
         year = 2017,
        month = feb,
       volume = {465},
       number = {2},
        pages = {1384-1400},
          doi = {10.1093/mnras/stw2766},
archivePrefix = {arXiv},
       eprint = {1610.06939},
 primaryClass = {astro-ph.GA},
       adsurl = {https://ui.adsabs.harvard.edu/abs/2017MNRAS.465.1384C}
}

@ARTICLE{andrews2013mass,
       author = {{Andrews}, Brett H. and {Martini}, Paul},
        title = "{The Mass-Metallicity Relation with the Direct Method on Stacked Spectra of SDSS Galaxies}",
      journal = {\apj},
         year = 2013,
        month = mar,
       volume = {765},
       number = {2},
          eid = {140},
        pages = {140},
          doi = {10.1088/0004-637X/765/2/140},
archivePrefix = {arXiv},
       eprint = {1211.3418},
 primaryClass = {astro-ph.CO},
       adsurl = {https://ui.adsabs.harvard.edu/abs/2013ApJ...765..140A}
}

@ARTICLE{luridiana2015pyneb,
       author = {{Luridiana}, V. and {Morisset}, C. and {Shaw}, R.~A.},
        title = "{PyNeb: a new tool for analyzing emission lines. I. Code description and validation of results}",
      journal = {\aap},
         year = 2015,
        month = jan,
       volume = {573},
          eid = {A42},
        pages = {A42},
          doi = {10.1051/0004-6361/201323152},
archivePrefix = {arXiv},
       eprint = {1410.6662},
 primaryClass = {astro-ph.IM},
       adsurl = {https://ui.adsabs.harvard.edu/abs/2015A&A...573A..42L}
}

@ARTICLE{vaught2024investigating,
       author = {{Rickards Vaught}, Ryan J. and {Sandstrom}, Karin M. and {Belfiore}, Francesco and {Kreckel}, Kathryn and {M{\'e}ndez-Delgado}, J. Eduardo and {Emsellem}, Eric and {Groves}, Brent and {Blanc}, Guillermo A. and {Dale}, Daniel A. and {Egorov}, Oleg V. and {Glover}, Simon C.~O. and {Grasha}, Kathryn and {Klessen}, Ralf S. and {Neumann}, Justus and {Williams}, Thomas G.},
        title = "{Investigating the Drivers of Electron Temperature Variations in H II Regions with Keck-KCWI and VLT-MUSE}",
      journal = {\apj},
         year = 2024,
        month = may,
       volume = {966},
       number = {1},
          eid = {130},
        pages = {130},
          doi = {10.3847/1538-4357/ad303c},
archivePrefix = {arXiv},
       eprint = {2309.17440},
 primaryClass = {astro-ph.GA},
       adsurl = {https://ui.adsabs.harvard.edu/abs/2024ApJ...966..130R}
}

@ARTICLE{ji2022correlation,
       author = {{Ji}, Xihan and {Yan}, Renbin},
        title = "{Correlation between the gas-phase metallicity and ionization parameter in extragalactic H II regions}",
      journal = {\aap},
         year = 2022,
        month = mar,
       volume = {659},
          eid = {A112},
        pages = {A112},
          doi = {10.1051/0004-6361/202142312},
archivePrefix = {arXiv},
       eprint = {2110.00612},
 primaryClass = {astro-ph.GA},
       adsurl = {https://ui.adsabs.harvard.edu/abs/2022A&A...659A.112J}
}

@ARTICLE{lee2024ionized,
       author = {{Lee}, Man-Yin Leo and {Yan}, Renbin and {Ji}, Xihan and {Algodon}, Gerome and {Westfall}, Kyle and {Lin}, Zesen and {Belfiore}, Francesco and {Bundy}, Kevin},
        title = "{Ionized gas in quiescent galaxies: Temperature measurement and constraint on the ionization source}",
      journal = {\aap},
         year = 2024,
        month = oct,
       volume = {690},
          eid = {A83},
        pages = {A83},
          doi = {10.1051/0004-6361/202348459},
archivePrefix = {arXiv},
       eprint = {2408.07952},
 primaryClass = {astro-ph.GA},
       adsurl = {https://ui.adsabs.harvard.edu/abs/2024A&A...690A..83L}
}
\bibliographystyle{aasjournalv7}



\end{document}